\documentclass[useAMS,usenatbib]{mn2e}

\usepackage{graphicx}

\title[IR photometry of YMCs in NGC 4214]{Infrared photometry of Young Massive Clusters in the
starburst galaxy NGC 4214\thanks{Based on NICS observations made with the Italian 
Telescopio Nazionale Galileo (TNG) operated on the island of La Palma by the 
Fundaci\'on Galileo Galilei of the INAF (Istituto Nazionale di Astrofisica) at 
the Spanish Observatorio del Roque de los Muchachos of the Instituto de 
Astrofisica de Canarias, within the observing program A25TAC\_2}}
\author[Sollima et al.]{A. Sollima$^{1,2}$\thanks{E-mail:
antonio.sollima@oapd.inaf.it}, R. G. Gratton$^{1}$, E. Carretta$^{2}$, A.
Bragaglia$^{2}$, S. Lucatello$^{1}$\\
$^{1}$ INAF Osservatorio Astronomico di Padova, vicolo dell'Osservatorio 5,
Padova, 35122, Italy\\
$^{2}$ INAF Osservatorio Astronomico di Bologna, via Ranzani 1, Bologna, 40127,
Italy}
\begin{document}


\pagerange{\pageref{firstpage}--\pageref{lastpage}} \pubyear{2013}

\maketitle

\label{firstpage}

\begin{abstract}
We present the results of an infrared photometric survey performed with 
NICS@TNG in the
nearby starburst galaxy NGC 4214. We derived accurate integrated JK magnitudes 
of 10 young massive clusters and compared them with the already available
Hubble Space Telescope ultraviolet colors. These clusters are located in
the combined ultraviolet-infrared colors planes on well defined sequences, whose
shapes allow a precise determination of their age. 
By means of the comparison with suitable stellar evolution 
models we estimated ages, metallicities,
reddening and masses of these clusters. All the analyzed clusters appear to be
younger than $\log{t/yr}<8.4$, moderately metal-rich and slightly less massive 
than present-day Galactic globular clusters. The derived ages for clusters
belonging to the secondary HII star forming complex are significantly 
larger than those previously estimated in the literature. We also discuss the
possibility of using the ultraviolet-infrared color-color diagram to select
candidate young massive clusters hosting multiple stellar populations.
\end{abstract}


\section{Introduction}
\label{intro_sec}

NGC 4214 is a nearby dwarf IAB(s)m galaxy (de Vaucouleurs et al. 1991) in the 
low-redshift CVn I Cloud (Sandage \& Bedke 1994) located at a distance of $3.04
\pm 0.05$ Mpc (Dalcanton et al. 2009). It has a mass similar to that of the
Large Magellanic Cloud (LMC; $M=5\times 10^{9}~M_{\odot}$; Karachentsev et al. 2004)
and a moderately low metal-content ($-1.6<[M/H]<0$; Williams et al. 2011
hereafter W11).
NGC 4214 is characterized by an intense recent star-formation activity, as shown
by the presence of two HII star forming complexes located in its central region
where several Young Massive Clusters (YMCs) are evident as H$\alpha$ emitting peaks 
(Fanelli et al. 1997). 
The young stellar
population is embedded in a smooth disk of old stars which account for a
significant fraction (up to $\sim 75\%$) of the total stellar mass (W11). 

\begin{figure}
 \includegraphics[width=8.7cm]{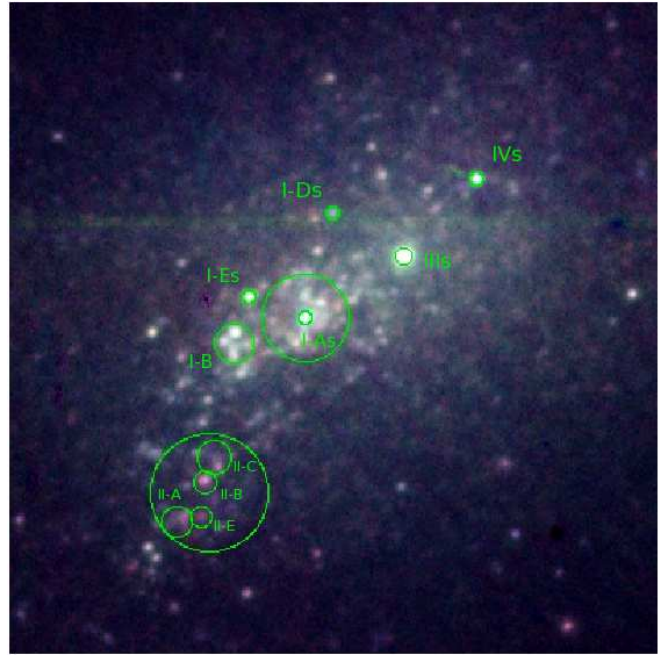}
\caption{Zoomed false-color image of NGC 4214 made using our NICS data. The J,H
and K images have been combined in the blue, green and red
channels, respectively (in the printed version of the paper the K band 
image is shown in greyscale). North is up, 
east to the left. The adopted circular apertures are shown with green circles. 
The field dimensions are $1.4\arcmin \times 1.4\arcmin$.}
\label{ima}
\end{figure}

MacKenty et al. (2000) provided a detailed classification of the
$H\alpha$ knots visible in Hubble Space Telescope (HST) images, 
identifying 13 relatively bright YMCs (see also Fig. \ref{ima}). 
The largest star forming complex is located in the north-western region of NGC
4214 (referred as 4214-I). It is formed by a number of star clusters and has a
complex $H\alpha$ structure dominated by the presence of two cavities. 
Its ultraviolet flux is dominated by an extended YMC (I-As)
extending for $\sim$ 70 pc located in a heart-shaped $H\alpha$ cavity. The
shell structure of the gas suggests that most of the gas in front of the central
star cluster has been removed by the action of stellar winds and supernovae
(see also Ma{\'{\i}}z-Apell{\'a}niz et al. 1998). The
other cavity is occupied by a scaled OB association (I-B) which, at odds with
the central cluster, does not show a marked central concentration. Other four
knots (I-C, I-Ds, I-Es and I-F) have been identified, probably associated to small
associations.
In the second complex (4214-II) there is no 
dominant star cluster but five smaller OB associations (II-A, II-B, II-C, II-D
and II-E), responsible for the
ionization of the gas, have been detected. In this region there is no clear evidence of 
decoupling between gas and stars. 
Both complexes present similar metallicities ($[M/H]\sim$-0.5; Kobulnicky \&
Skillman 1996) with little dispersion.
Other two clusters (IIs and IVs) are present outside the two main complexes, 
both characterized by a weak $H\alpha$ emission and redder colors.
Dust absorption appears to be on average relatively low ($E(B-V)<0.1$) although patchy
with some regions heavily obscured (Ma{\'{\i}}z-Apell{\'a}niz 1998; Drozdovsky et al. 2002;
Calzetti et al. 2004; {\'U}beda, Ma{\'{\i}}z-Apell{\'a}niz \& MacKenty 2007a; hereafter U07a).

While the diffuse stellar population of NGC 4214 has been the subject of many
recent works (Drozdovsky et al. 2002; W11), the study of 
the star cluster system of NGC 4214 has been restricted to the integrated
properties of the two main complexes 
(Leitherer et al. 1996; Mas-Hesse \& Kunth 1999; Hermelo et al. 2012).
Recently {\'U}beda et al. (2007a,b) performed an accurate study using 
WFPC2@HST observations associated
to 2MASS infrared photometry to determine ages, masses, radii and extinction of 
11 YMCs. The stellar extinction they estimated appears to follow a
Magellanic Cloud-like law and has been found to be quite 
patchy, with some heavily obscured areas, in particular
those associated with star-forming regions. The clusters in complex I and II
have been estimated to be very young ($\log{t/yr}<7$) while clusters IIIs and IVs
show significantly older ages ($\log{t/yr}\sim 8$).

The study of the stellar population of YMCs has deep relevance in the light of 
the scenario of formation of multiple stellar populations in globular clusters. 
In fact, in the last decade a growing body of evidence showed that 
many Galactic globular clusters contain multiple stellar
populations which have been evidenced as multiple sequences in various 
regions of the color-magnitude diagram
(e.g. Piotto 2009) and/or as distinct groups of stars
with peculiar chemical abundances (Gratton, Carretta \& Bragaglia 2012).
The above evidence suggests that the stellar populations of these 
objects are the result of a process of self-enrichment where a second
generation of stars formed from the ejecta of a first generation. 
The observed pattern of abundance variations (light elements
anticorrelations with almost constant Fe abundance; Carretta et al. 2009a,b,c) 
indicates that the
responsibles of the chemical pollution must be massive
($M>5~M_{\odot}$) stars undergoing high-temperature p-capture nuclear
reactions but not supernovae (Langer, Hoffman \& Sneden 1993). 
As a consequence, the entire process of chemical enrichment must have taken
place within the first $10^8$ yr i.e. the timescale of evolution of such massive
stars (see e.g. D'Antona \& Ventura 2007; Decressin et al. 2007 for two possible
alternative hypotheses on the nature of the polluters).
To date, the phenomenon of multiple populations has been observed in almost all 
the analyzed globular clusters (with masses $M>10^{5}M_{\odot}$), but not 
in less massive systems like e.g. open clusters
(de Silva et al. 2009; Bragaglia et al. 2012; Geisler et al. 2012). This seems to indicate that the process of formation
of multiple populations could start only in systems above a critical mass, 
able to retain the gas expelled by the early generation of massive stars 
(Carretta et al. 2010; Bekki 2011).
YMCs represent therefore a particularly favorable class of objects to study this 
phenomenon (see also Peacock, Zempf \& Finzell 2013). While resolved photometry is becoming feasible for YMCs in
nearby galaxies (Perina et al. 2010; Larsen et al. 2011) the large majority
of YMCs are located in distant galaxies where only their integrated
properties can be studied.

In this paper we present infrared JK photometry of 10 YMCs of NGC 4214 obtained
using NICS@TNG. These data together with the ultraviolet/optical HST 
photometry of {\'U}beda et al. (2007b; hereafter U07b) are used to improve the ages,
metallicities, masses and reddening determinations of these YMCs. 
The paper is organized as follow: in Sect. 2 we describe the observations and the data
reduction technique. Sect. 3 is devoted to the analysis of the location of YMCs
in the ultraviolet-infrared colors planes and the comparison with model 
predictions. In Sect. 4 the method to derive ages, metallicities and
reddening is outlined and the results presented.  
In Sect. 5 the potential application to multiple stellar populations studies is
discussed.
Finally, we summarize and discuss our results in Sect. 6.

\section{Observations and Data Reduction} 
\label{obs_sec}

\begin{table*}
 \centering
 \begin{minipage}{140mm}
  \caption{JK magnitudes of YMCs in NGC 4214.}
  \begin{tabular}{@{}lccccccr@{}}
  \hline
   cluster & RA (J2000) & Dec (J2000) & r          & J & $\sigma_{J}$ & K & $\sigma_{K}$\\
           & deg        & deg         & $\arcsec$  &   &              &   &\\
 \hline
  I-As &  183.9147500 &  36.3268028 & 1.00 & 15.889 & 0.016 & 15.673  & 0.026\\
  I-A  &  183.9147083 &  36.3269444 & 6.83 & 13.775 & 0.030 & 13.120  & 0.026\\
  I-B  &  183.9183333 &  36.3258833 & 2.96 & 14.896 & 0.008 & 14.129  & 0.018\\
  I-Ds &  183.9134167 &  36.3313778 & 1.00 & 17.400 & 0.017 & 17.128  & 0.053\\
  I-Es &  183.9177500 &  36.3277611 & 1.00 & 16.708 & 0.023 & 15.881  & 0.018\\
  II-A &  183.9216667 &  36.3180889 & 2.28 & 16.424 & 0.014 & 15.538  & 0.029\\
  II-B &  183.9198750 &  36.3197389 & 1.68 & 16.783 & 0.017 & 15.631  & 0.010\\
  II-C &  183.9196667 &  36.3208639 & 2.59 & 16.604 & 0.019 & 15.773  & 0.033\\
  II-E &  183.9202917 &  36.3179750 & 1.59 & 17.549 & 0.020 & 16.497  & 0.035\\
   II  &  183.9199583 &  36.3192389 & 9.10 & 14.382 & 0.036 & 13.505  & 0.022\\
  IIIs &  183.9094583 &  36.3295250 & 1.20 & 14.962 & 0.013 & 14.231  & 0.014\\
   IVs &  183.9055833 &  36.3328333 & 1.00 & 16.427 & 0.014 & 15.893  & 0.024\\
\hline
\end{tabular}
\end{minipage}
\end{table*}

Observations were performed during one night on April 11th 2012 at the 
Telescopio Nazionale Galileo (TNG; 
Roche de los Muchachos, Spain), equipped with the Near Infrared Camera Spectrometer
(NICS). Observations were performed in Service Mode using the large field camera
providing a field of view of $4.2\arcmin \times 4.2\arcmin$ with a pixel scale of
$0.25\arcsec /px$. The seeing conditions were stable during observations with a
$FWHM\sim1.0\arcsec$.
A set of 9 images with exposure time $3\times20s$
($NDIT\times DIT$) has been taken through the J filter and 15 images with
exposure time $1\times60s$ through the K one, all centered on
NGC 4214. Images in the H band were also observed but their signal-to-noise
ratios were significantly lower than those in the J and K bands and 
were not used in the analysis. Science fields have been observed alternately with sky fields (at 
$\sim 12$ arcmin away from the science field center in a square pattern with 
the same exposure times of science fields).
A median sky image was obtained for each filter and subtracted to science frames. 
A set of high-S/N flat fields in each filter has been obtained with a halogen 
lamp and has been used to correct sky-subtracted frames.

We used the aperture photometry algorithm PHOT of the {\rm DAOPHOT II} package 
(Stetson 1987) to obtain instrumental magnitudes for all the 10 YMCs detected 
in each frame (I-As, I-B, I-Ds, I-Es, II-A, II-B, II-C, II-E, IIIs, IVs). 
The photometry has been performed independently on each frame and calibrated to
the 2MASS photometric system using four bright stars present in the field of 
view of our science frame. The final magnitudes have been
obtained as the average of single exposure measures and the related r.m.s. has
been assigned as their error.
We converted positions from the detector reference frame to celestial
coordinates using $\sim$100 stars well distributed over the NICS field of view and
present in the 2MASS catalog. The accuracy of the astrometric solution turns out
to be $\sim0.2\arcsec$.
We adopted the same apertures defined by U07b for most clusters, 
with the exceptions of clusters I-As, I-Ds, I-Es and IVs: for these objects the
apertures defined by U07b are smaller than the resolution of our data, so we
adopted in these cases an aperture radius of $1\arcsec$ (see Fig. \ref{ima}).
For this reason, near infrared JK and HST magnitudes do not
contain the same fraction of the cluster flux and therefore cannot be 
directly compared. The same considerations can be made for the other
clusters: as the spatial resolution of our J and K images is significantly 
worse than that of HST observations, a small but
significant part of the cluster light could be spread out the adopted
aperture in infrared bands. Thus, only the J-K colors (virtually independent on
the aperture size) have been used in the present analysis.
As in U07b, two large apertures have been also adopted to measure the integrated
magnitudes of the two main star forming complexes (IA and II).
The obtained JK magnitudes are listed in Table 1.

The derived JK magnitudes have been complemented with the ultraviolet-optical 
magnitudes obtained by U07b using the WFPC2 onboard HST. We adopted the
magnitudes listed in their Table 6 and 7 which, beside the basic data reduction
steps, account for an aperture correction to link the flux measured within the
aperture to that at infinity.

\section{The ultraviolet-infrared color-color diagrams}
\label{uvir_sec}

\begin{figure*}
 \includegraphics[width=15cm]{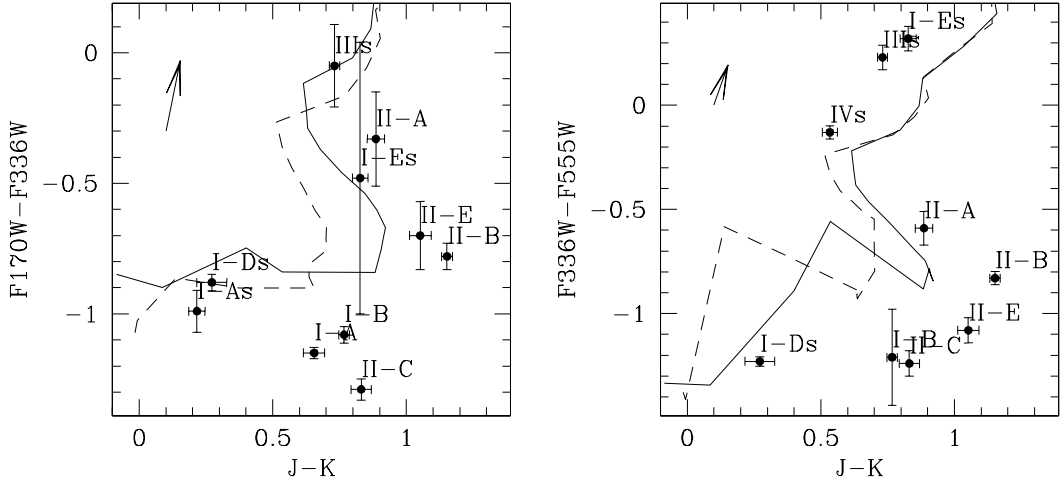}
\caption{Position of the YMCs of NGC 4214 in the $(F170W-F336W)~vs.~(J-K)$ (left
panel) and $(F336W-F555W) vs. (J-K)$ (right panel) color-color diagrams.
The reddening vectors for E(B-V)=0.1 are shown. 
Evolutionary models corresponding to metallicities
$[M/H]=-0.5$ (dashed lines) and
$[M/H]=0$ (solid lines) are overplotted in both panels.}
\label{colcol}
\end{figure*}

In Fig. \ref{colcol} the location of the observed clusters in the 
$(F170W-F336W) vs. (J-K)$ and $(F336W-F555W) vs. (J-K)$
color-color diagrams is shown. It can be noted that in both diagrams the YMCs of
NGC 4214 align along well defined, non-monotonic relations: as the ultraviolet colors
get redder the J-K color spans a wide range with a first increase (up to
$J-K\sim1.2$ at $F336W-F555W\sim-0.8$), a subsequent blue shift ($J-K\sim0.5$ at
$F336W-F555W\sim-0.2$) and a final reddening.
The theoretical evolution in these planes of a young cluster adopting two 
different metallicities ($[M/H]=-0.5,0$; appropriated for the recent star
formation episodes of a dwarf galaxy such as NGC 4214; W11) are overplotted in both 
diagrams.
These predictions have been obtained by integrating the fluxes predicted by a set of 
Marigo et al. (2008) isochrones adopting the Kroupa (2001) mass function.
We note that, while models fail to reproduce the precise extent of the
clusters distribution in the ultraviolet-infrared color-color diagrams, they qualitatively 
predict their characteristic S-shapes.
In fact, stellar population models indicate
that during the early evolution of a single stellar population, while
ultraviolet colors (dominated by the contribution of Main Sequence stars) 
become steadily redder, infrared colors are
characterized by an initial shift toward the red (after $\sim6~10^{6}~yr$ due to
the formation of a significant number of Red Supergiants) and a subsequent
turnover at $10^{8}~yr$ (when the onset of He burning moves stars less massive
than $8-9~M_{\odot}$ toward blue colors). At ages larger than $10^{8}$ yr the infrared colors become
again redder due to the onset of the extended Asymptotic Giant Branch (AGB) 
phase in stars with masses $M<5~M_{\odot}$.
The narrow sequence defined by our data in these planes suggests a rather
homogeneous metallicity and reddening distribution across the galaxy. 
The peculiar shape of the ultraviolet-infrared color-color diagrams allows a
very precise determination of clusters ages which is less dependent on
the temperature-colors conversion uncertainties than analyses based on optical 
colors alone (e.g. Fall, Chandar \& Whitmore 2005). 
Indeed, as the infrared color turnovers correspond to 
phase transitions occurring during the evolution of massive stars, it is
possible to infer ages according to the relative position of clusters with
respect to the {\it observed} position of such turnovers. The potential power of infrared colors in determining ages of young clusters has
been also recently reported by Gazak et al. (2012) who applied a similar
technique to a sample of young clusters in M83.
Optical colors are
less efficient in this regard since they are sensitive to both Main Sequence
and evolved stars producing colors always monotonically correlated between them.

\subsection{Comparison with LMC clusters and theoretical models}
\label{lmc_sec}

\begin{figure}
 \includegraphics[width=8.7cm]{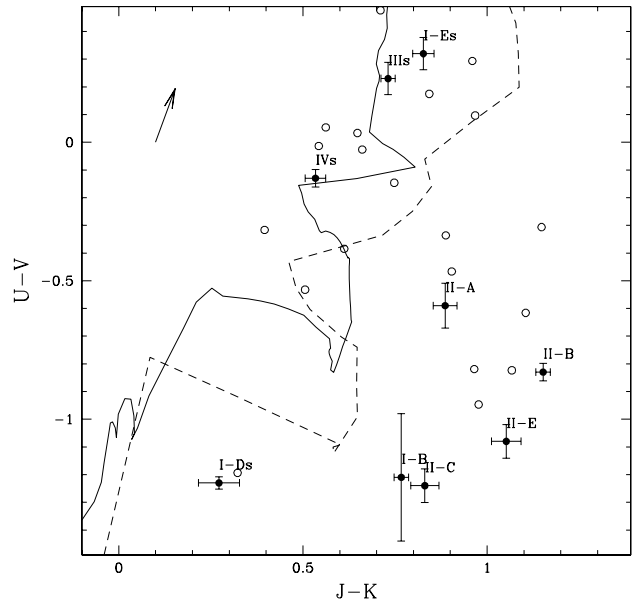}
\caption{Position in the $(U-V)~vs.~(J-K)$ color-color diagram of the 
YMCs of NGC 4214 (filled circles) and of the LMC (from Searle et al. 1980 and Kyeong
et al. 2003; open circles).
The reddening vector for E(B-V)=0.1 is shown. 
The evolutionary tracks corresponding to a metallicity 
$[M/H]=-0.5$ adopting the stellar population synthesis procedure described in this paper 
(dashed line) and the 
Bruzual \& Charlot (2003) models (solid line) are overplotted.}
\label{lmc}
\end{figure}

\begin{figure}
 \includegraphics[width=8.7cm]{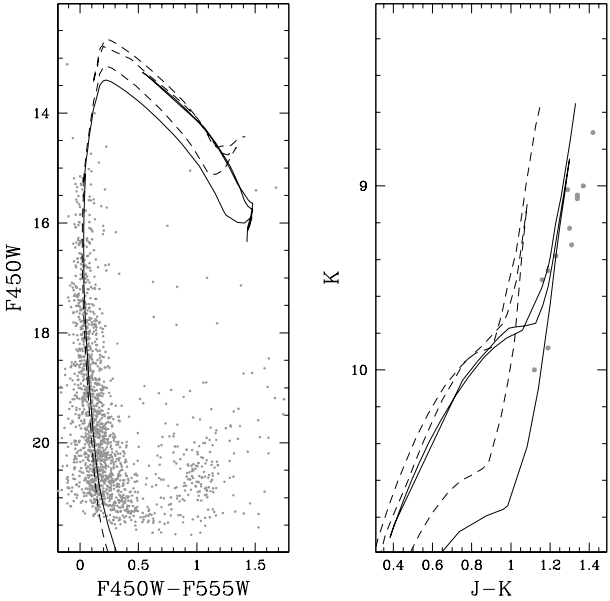}
\caption{$F450W - (F450W-F555W)$ (from Brocato et al. 2001; left panel) and 
$K -(J-K)$ (from Keller 1999; right panel) CMDs of NGC 2100. 
The isochrones corresponding to a metallicity 
$[M/H]=-0.5$ (dashed line) and $[M/H]=0$ (solid lines) are overplotted in both
panels.}
\label{ngc2100}
\end{figure}

As outlined in Sect. \ref{intro_sec}, NGC 4214 and the LMC share many
similarities: they have the same morphological type, similar masses
and metal content and both possess a significant population of YMCs.
It is therefore interesting to compare the location of their YMCs in the 
ultraviolet-infrared color-color plane.
This is shown in Fig. \ref{lmc} where the U-V (fairly similar to the
$F336W-F555W$ color) and J-K colors for LMC clusters
were taken from Searle, Wilkinson \& Bagnuolo (1980) and Kyeong et al.
(2003), respectively. 
It is evident that LMC clusters nicely follow the sequence defined by those in
NGC 4214, although with a larger scatter. 

In both cases stellar
populations models provide a poor representation of the data, predicting a
significantly less pronounced red excursion at $U-V\sim-0.8$ during the Red
Supergiants phase. The same effect is evident adopting different stellar
population synthesis models (like Bruzual \& Charlot 2003\footnote{Both the
models adopted in the present work and those by Bruzual \& Charlot (2003) make
use of the isochrone set by the Padova group. A similar discrepancy
is also found using other popular models (see V{\'a}zquez \& Leitherer 2005).}; see Fig. \ref{lmc}). 
While errors on the photometric calibration and
small number fluctuations can contribute to such a discrepancy, its presence
in both galaxies suggests a problem in the model predictions.
The reason of the inadequacy of models can be found by comparing the adopted
isochrones with the resolved photometric data available for some LMC cluster
populating this region of the $U-V vs. (J-K)$ color-color diagram.
In Fig. \ref{ngc2100} the $F450W - (F450W-F555W)$ (from Brocato, Di Carlo \&
Menna 2001) 
and $K - (J-K)$ (from Keller 1999)
color-magnitude diagrams (CMDs) of the cluster NGC~2100 
(having integrated colors $U-V=-0.4$ and $J-K=1.19$) are compared with the 
same isochrones of the Marigo et al. (2008) set used to compute the stellar 
population evolutionary sequences in Fig. \ref{colcol} (with metallicity 
$[M/H]=-0.5$ and $[M/H]=0$, bracketing the metallicity estimates for this 
cluster by Jasniewicz \& Thevenin
1994, Hill \& Spite 1999 and Colucci et al. 2011). Adopting a distance 
modulus of $(m-M)_{0}=18.50$
(Sebo et al. 2002) the bestfit age and reddening for this cluster 
turn out to be $\log{t/yr}=7.35$ and $E(B-V)=0.3$, almost independent on the 
adopted metallicity. The isochrones, while fitting well the 
optical CMD, are some 0.1-0.3 mag bluer than the observed stars in
the infrared CMD. 
Such a difference can account entirely for the J-K color difference between 
models and observations in Figs. \ref{colcol} and \ref{lmc} and
is probably linked to the uncertainties in the color-temperature
transformations for the cold ($T\sim3500~K$) Red Supergiants which dominate the
infrared luminosity. 

However, although stellar evolution models do not provide an
optimal representation of the data in the age interval 
6.8 $< \log{t/yr}<$ 8 (where most of the considered clusters lie), they 
qualitatively predict the shift toward red colors observed in both NGC 4214 and 
LMC YMCs. This allows to identify those clusters in the relatively fast stage 
in which Red Supergiants dominate their infrared flux, whose ages could not
be unambiguously determined using ultraviolet/optical colors alone. 
So, in spite of the apparent discrepancy between models and
observations, the addition of the IR dimension in the parameter space is
still valuable.

\section{Age determination}

\begin{table*}
 \centering
 \begin{minipage}{140mm}
  \caption{Derived parameters of YMCs in NGC 4214.}
  \begin{tabular}{@{}lcccccccccr@{}}
  \hline
  & \multicolumn{10}{c}{Metallicity within $\Delta[M/H]<-0.5$}\\
  \hline
  & \multicolumn{5}{c}{Reddening left free} & \multicolumn{5}{c}{Reddening within
  $\Delta E(B-V)<0.1$}\\
   cluster & $\log{t/yr}$ & [M/H] & $\log{M/M_{\odot}}$ & E(B-V) & $\chi^{2}/n$ & $\log{t/yr}$ & 
   [M/H] & $\log{M/M_{\odot}}$ & E(B-V) & $\chi^{2}/n$\\
 \hline
  I-As & 6.8 & -0.3 & 4.81 & 0.04 & 0.005 & 6.8 & -0.3 & 4.81 & 0.04 & 0.005\\
       & ($\pm$0.1) & ($\pm$0.1) & ($\pm$0.02) & ($\pm$0.01) &  & ($\pm$0.1) & ($\pm$0.1) & ($\pm$0.02) & ($\pm$0.01) &\\
  I-A  & 7.0 & -0.4 & 5.61 & 0.00 & 0.008 & 7.0 & -0.3 & 5.68 & 0.03 & 0.712\\
       & ($\pm$0.1) & ($\pm$0.1) & ($\pm$0.02) & ($\pm$0.01) &  & ($\pm$0.1) & ($\pm$0.3) & ($\pm$0.06) & ($\pm$0.01) &\\
  I-B  & 7.0 & -0.1 & 4.99 & 0.01 & 0.135 & 7.0 & -0.2 & 5.03 & 0.03 & 0.210\\
       & ($\pm$0.2) & ($\pm$0.2) & ($\pm$0.04) & ($\pm$0.02) &  & ($\pm$0.1) & ($\pm$0.3) & ($\pm$0.05) & ($\pm$0.01) &\\
  I-Ds & 6.7 &  0.0 & 4.15 & 0.13 & 1.199 & 6.7 &  0.0 & 4.15 & 0.13 & 1.199\\
       & ($\pm$0.1) & ($\pm$0.5) & ($\pm$0.12) & ($\pm$0.09) &  & ($\pm$0.1) & ($\pm$0.3) & ($\pm$0.12) & ($\pm$0.05) &\\
  I-Es & 6.9 &  0.0 & 4.07 & 0.54 & 0.771 & 8.3 &  0.2 & 4.68 & 0.13 & 1.002\\
       & ($\pm$0.1) & ($\pm$0.3) & ($\pm$0.10) & ($\pm$0.12) &  & ($\pm$0.2) & ($\pm$0.4) & ($\pm$0.11) & ($\pm$0.05) &\\
    "  & 8.0 & -0.5 & 4.70 & 0.31 & 0.828 &     &      &      &      &      \\
       & ($\pm$0.4) & ($\pm$0.3) & ($\pm$0.08) & ($\pm$0.07) &  &            &            &             &             &\\
  II-A & 7.2 & -0.2 & 5.17 & 0.25 & 0.013 & 7.5 &  0.2 & 5.21 & 0.13 & 0.101\\
       & ($\pm$0.2) & ($\pm$0.1) & ($\pm$0.07) & ($\pm$0.05) &  & ($\pm$0.1) & ($\pm$0.3) & ($\pm$0.03) & ($\pm$0.02) &\\
  II-B & 7.2 &  0.0 & 4.80 & 0.09 & 2.852 & 7.2 &  0.0 & 4.80 & 0.09 & 2.852\\
       & ($\pm$0.3) & ($\pm$0.3) & ($\pm$0.15) & ($\pm$0.06) &  & ($\pm$0.3) & ($\pm$0.3) & ($\pm$0.15) & ($\pm$0.05) &\\
  II-C & 7.0 & -0.2 & 4.61 & 0.00 & 2.800 & 7.0 & -0.3 & 4.66 & 0.03 & 5.437\\
       & ($\pm$0.1) & ($\pm$0.3) & ($\pm$0.21) & ($\pm$0.01) &  & ($\pm$0.2) & ($\pm$0.5) & ($\pm$0.35) & ($\pm$0.01) &\\
  II-E & 7.3 &  0.0 & 4.53 & 0.03 & 1.762 & 7.3 &  0.1 & 4.52 & 0.03 & 1.757\\
       & ($\pm$0.3) & ($\pm$0.3) & ($\pm$0.17) & ($\pm$0.05) &  & ($\pm$0.3) & ($\pm$0.3) & ($\pm$0.11) & ($\pm$0.03) &\\
   II  & 7.1 &  0.0 & 5.52 & 0.01 & 2.599 & 7.1 &  0.0 & 5.56 & 0.03 & 2.893\\
       & ($\pm$0.3) & ($\pm$0.5) & ($\pm$0.39) & ($\pm$0.03) &  & ($\pm$0.3) & ($\pm$0.3) & ($\pm$0.22) & ($\pm$0.02) &\\
  IIIs & 7.7 & -0.2 & 5.23 & 0.24 & 0.001 & 7.9 &  0.2 & 5.23 & 0.13 & 0.010\\
       & ($\pm$0.1) & ($\pm$0.1) & ($\pm$0.03) & ($\pm$0.01) &  & ($\pm$0.1) & ($\pm$0.1) & ($\pm$0.15) & ($\pm$0.01) &\\
   IVs & 6.8 & -0.1 & 4.56 & 0.46 & 0.002 & 8.0 & -0.3 & 5.13 & 0.13 & 0.217\\
       & ($\pm$0.1) & ($\pm$0.1) & ($\pm$0.05) & ($\pm$0.03) &  & ($\pm$0.1) & ($\pm$0.2) & ($\pm$0.06) & ($\pm$0.01) &\\
    "  & 8.0 & -0.5 & 5.13 & 0.15 & 0.006 &     &      &      &      &      \\
       & ($\pm$0.1) & ($\pm$0.1) & ($\pm$0.01) & ($\pm$0.01) &  &            &            &             &             &\\
\hline
\end{tabular}
\end{minipage}
\end{table*}

\begin{table*}
\addtocounter{table}{-1}
 \centering
 \begin{minipage}{140mm}
  \caption{Derived parameters of YMCs in NGC 4214. (continued)}
  \begin{tabular}{@{}lcccccccccr@{}}
  \hline
  & \multicolumn{10}{c}{Metallicity fixed to [M/H]=-0.5}\\
  \hline
  & \multicolumn{5}{c}{Reddening left free} & \multicolumn{5}{c}{Reddening within
  $\Delta E(B-V)<0.1$}\\
   cluster & $\log{t/yr}$ & [M/H] & $\log{M/M_{\odot}}$ & E(B-V) & $\chi^{2}/n$ & $\log{t/yr}$ & 
   [M/H] & $\log{M/M_{\odot}}$ & E(B-V) & $\chi^{2}/n$\\
 \hline
  I-As & 6.8 & -0.5 & 4.89 & 0.07 & 0.558 & 6.8 & -0.5 & 4.89 & 0.07 & 0.558\\
       & ($\pm$0.1) &  & ($\pm$0.08)  & ($\pm$0.03) &  & ($\pm$0.1) & & ($\pm$0.06) & ($\pm$0.03) &\\
  I-A  & 7.0 & -0.5 & 5.62 & 0.01 & 0.121 & 7.0 & -0.5 & 5.66 & 0.03 & 0.363\\
       & ($\pm$0.1) &  & ($\pm$0.02) & ($\pm$0.01) &  & ($\pm$0.1) & & ($\pm$0.06) & ($\pm$0.01) &\\
  I-B  & 7.3 & -0.5 & 5.23 & 0.00 & 1.071 & 7.0 & -0.5 & 5.01 & 0.04 & 1.157\\
       & ($\pm$0.3) &  & ($\pm$0.14) & ($\pm$0.02) &  & ($\pm$0.1) & & ($\pm$0.05) & ($\pm$0.01) &\\
  I-Ds & 6.6 & -0.5 & 4.17 & 0.18 & 1.899 & 6.7 & -0.5 & 4.13 & 0.13 & 3.679\\
       & ($\pm$0.1) & & ($\pm$0.13) & ($\pm$0.08) &  & ($\pm$0.1) & & ($\pm$0.12) & ($\pm$0.05) &\\
  I-Es & 7.1 & -0.5 & 4.55 & 0.68 & 2.129 & 8.4 & -0.5 & 4.63 & 0.13 & 1.189\\
       & ($\pm$0.2) & & ($\pm$0.28) & ($\pm$0.14) &  & ($\pm$0.2) & & ($\pm$0.12) & ($\pm$0.06) &\\
    "  & 8.1 & -0.5 & 4.70 & 0.31 & 0.904 &     &      &      &      &      \\
       & ($\pm$0.4) & & ($\pm$0.08) & ($\pm$0.07) &  &            &            &             &             &\\
  II-A & 7.3 & -0.5 & 5.22 & 0.24 & 0.631 & 7.4 & -0.5 & 5.10 & 0.13 & 1.880\\
       & ($\pm$0.2) & & ($\pm$0.08) & ($\pm$0.06) &  & ($\pm$0.1) & & ($\pm$0.024) & ($\pm$0.03) &\\
  II-B & 7.3 & -0.5 & 4.86 & 0.10 & 9.544 & 7.3 & -0.5 & 4.86 & 0.10 & 9.544\\
       & ($\pm$0.3) & & ($\pm$0.16) & ($\pm$0.07) &  & ($\pm$0.3) & & ($\pm$0.16) & ($\pm$0.06) &\\
  II-C & 7.0 & -0.5 & 4.57 & 0.00 & 3.801 & 7.0 & -0.5 & 4.64 & 0.03 & 5.915\\
       & ($\pm$0.1) & & ($\pm$0.21) & ($\pm$0.01) &  & ($\pm$0.2) & & ($\pm$0.35) & ($\pm$0.01) &\\
  II-E & 7.3 & -0.5 & 4.51 & 0.05 & 7.163 & 7.3 & -0.5 & 4.51 & 0.05 & 7.163\\
       & ($\pm$0.3) & & ($\pm$0.17) & ($\pm$0.05) &  & ($\pm$0.3) & & ($\pm$0.17) & ($\pm$0.05) &\\
   II  & 7.4 & -0.5 & 5.75 & 0.00 & 3.727 & 7.3 & -0.5 & 5.71 & 0.03 & 4.678\\
       & ($\pm$0.3) & & ($\pm$0.41) & ($\pm$0.04) &  & ($\pm$0.3) & & ($\pm$0.23) & ($\pm$0.03) &\\
  IIIs & 7.7 & -0.5 & 5.29 & 0.28 & 0.024 & 8.1 & -0.5 & 5.29 & 0.12 & 0.194\\
       & ($\pm$0.1) & & ($\pm$0.07) & ($\pm$0.04) &  & ($\pm$0.2) & & ($\pm$0.16) & ($\pm$0.02) &\\
   IVs & 6.9 & -0.5 & 4.61 & 0.41 & 0.006 & 8.0 & -0.5 & 5.11 & 0.13 & 0.189\\
       & ($\pm$0.1) & & ($\pm$0.06) & ($\pm$0.04) &  & ($\pm$0.1) & & ($\pm$0.06) & ($\pm$0.01) &\\
    "  & 8.0 & -0.5 & 5.13 & 0.15 & 0.010 &     &      &      &      &      \\
       & ($\pm$0.1) & & ($\pm$0.01) & ($\pm$0.01) &  &            &            &             &             &\\
\hline
\end{tabular}
\end{minipage}
\end{table*}

\begin{figure*}
 \includegraphics[width=12cm]{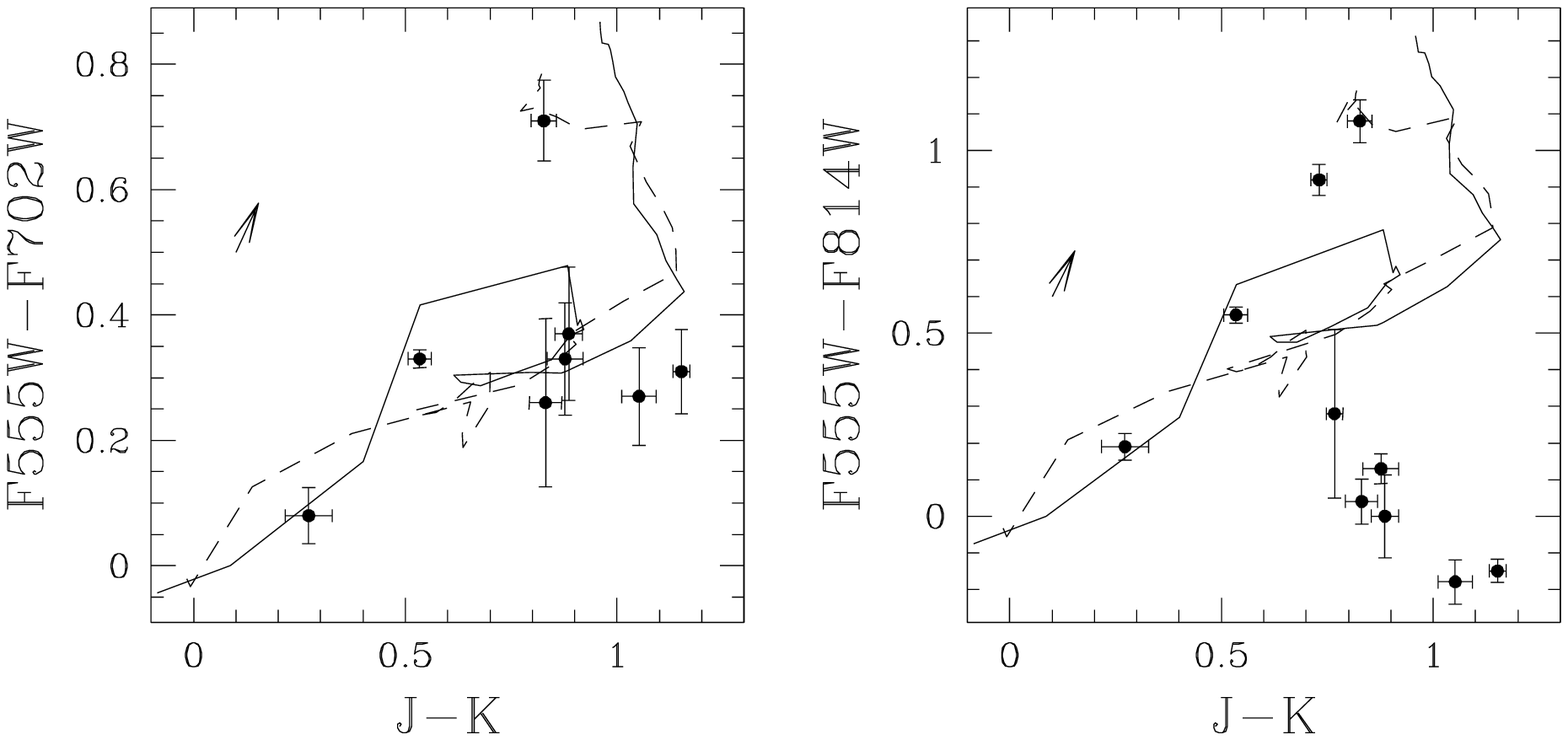}
\caption{Position of the YMCs of NGC 4214 in the $(F555W-F702W)~vs.~(J-K)$ (left
panel) and $(F555W-F814W) vs. (J-K)$ (right panel) color-color diagrams.
The reddening vectors for E(B-V)=0.1 are shown. 
Evolutionary models corresponding to metallicities
$[M/H]=-0.5$ (dashed lines) and
$[M/H]=0$ (solid lines) are overplotted in both panels.} 
\label{colex}
\end{figure*}

\begin{figure*}
 \includegraphics[width=12cm]{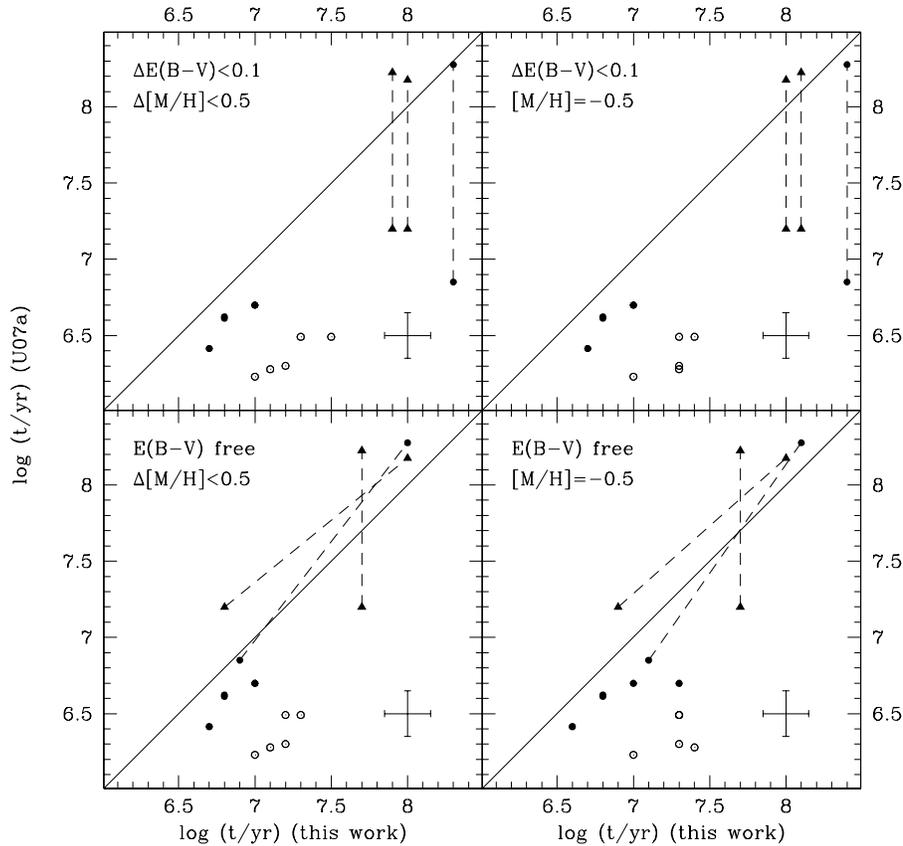}
\caption{Comparison between the ages 
estimated in this work and in U07a. In the bottom panels the comparison is made
with ages derived leaving the reddening as a free parameter,
while in the upper panels ages derived constraining the reddening to
span a $\Delta E(B-V)<0.1$ range are considered. In the left panels the comparison is made
with ages derived constraining the metallicity to
span a $\Delta [M/H]<0.5$ range, while in right panels ages derived assuming a fixed
metallicity [M/H]=-0.5 are considered. Filled dots indicate clusters belonging to
complex I, open dots indicate clusters belonging to complex II, triangles indicates
other clusters. The ranges covered by multiple age 
solutions are indicated by dashed lines. The one-to-one
relation is also shown as solid line in all panels. Typical
uncertainties are also indicated.} 
\label{confrt}
\end{figure*}

\begin{figure*}
 \includegraphics[width=12cm]{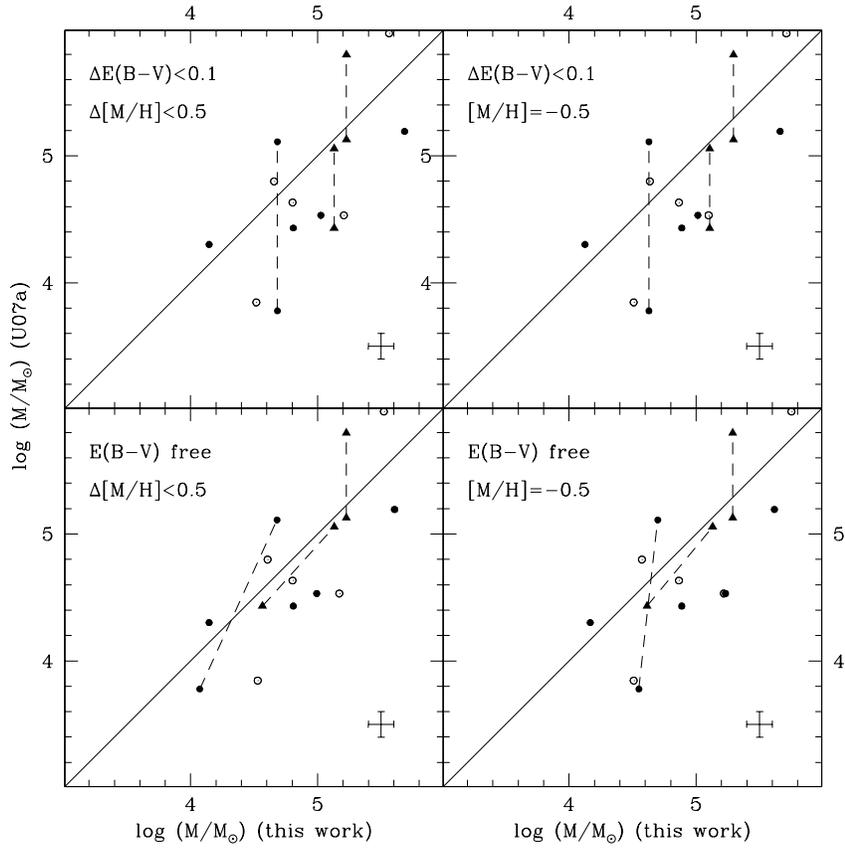}
\caption{Comparison between the masses 
estimated in this work and in U07a. Symbols and lines are as in Fig. \ref{confrt}.} 
\label{confrm}
\end{figure*}

\begin{figure*}
 \includegraphics[width=12cm]{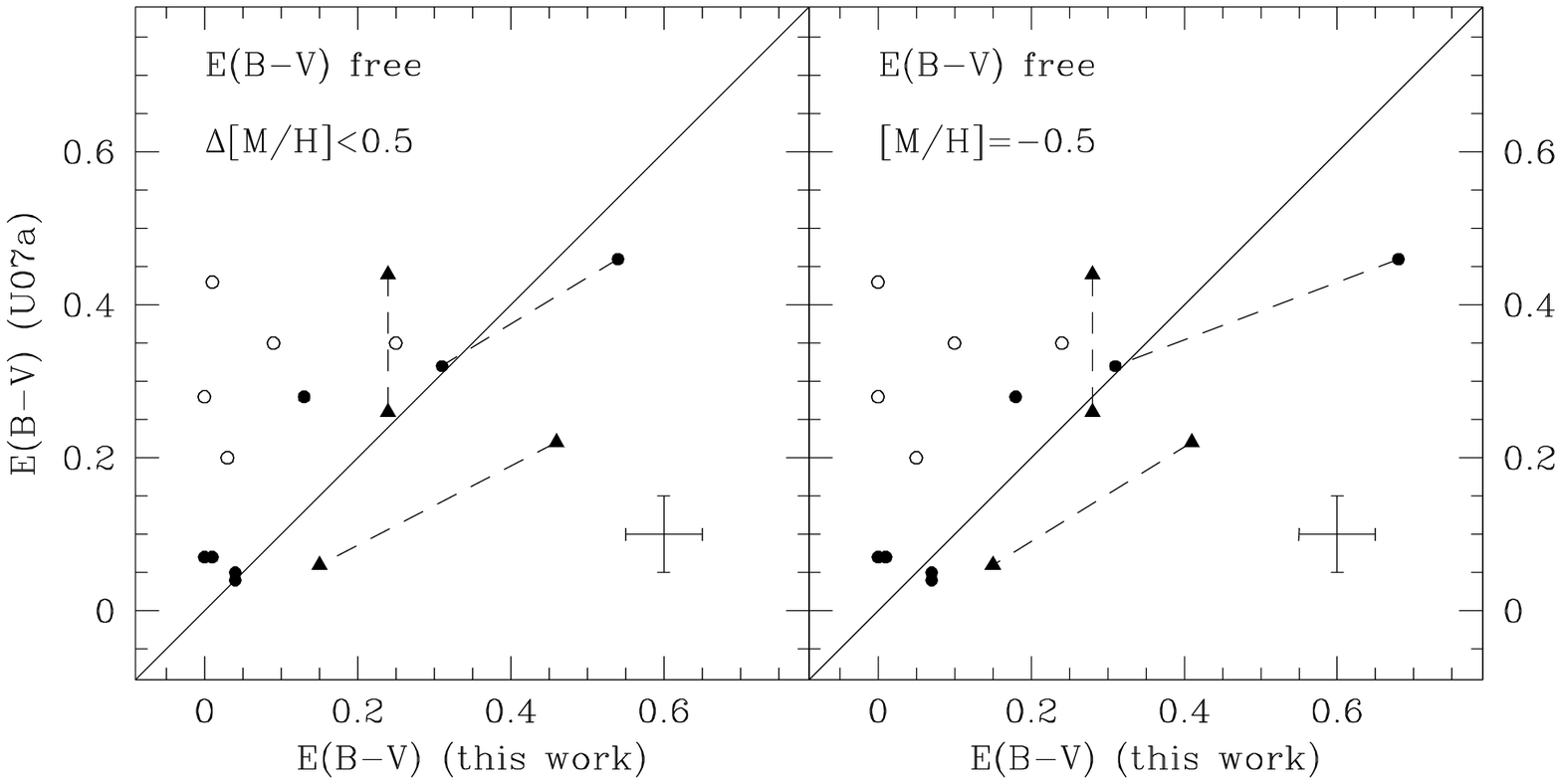}
\caption{Comparison between the reddening
estimated in this work and in U07a. Symbols and lines are as in Fig. 
\ref{confrt}.} 
\label{confrebv}
\end{figure*}

The accurate JK magnitudes, together with the ultraviolet/optical ones 
measured by U07b, have been used 
to derive the ages (t), metallicities ($[M/H]$), reddening ($E(B-V)$) and masses
(M) of the 10 observed YMCs using a $\chi^{2}$ minimization technique.
As outlined in Sect. \ref{obs_sec}, because of the different spatial
resolution between infrared data and HST observations, the apertures adopted
to derive JK magnitudes could contain a different fraction of the cluster flux
with respect to HST observations. In
this situation, our JK and HST magnitudes cannot be used together to search for the 
best model.
On the other hand, in absence of color 
gradients, the J-K color should be independent on the aperture size within the region 
dominated by the cluster.
So, we calculated the $\chi^{2}$ comparing each HST
magnitude and the J-K color with models (avoiding the individual
use of J and K magnitudes).
To check the dependence of the obtained results on this last choice, we 
repeated the analysis considering J and K 
magnitudes separately (instead of the J-K color alone): for
clusters with the same aperture the results remain unchanged (within 
$\Delta \log{t/yr}<$0.1) and found a discrepancy which grows almost linearly 
with the difference of aperture to a maximum of $\Delta \log{t/yr}\sim$0.3. So 
the above described effect is small while still present. Consider that even for 
those clusters for whom we adopted the same aperture of
U07a we cannot assume that HST and IR magnitudes contain the same fraction of
the cluster flux. For this reason we prefer to leave the J-K color as a single
independent constraint.

For each cluster, we searched the best combination of the above parameters that
minimize the penalty function\footnote{The formalism of eq. 1 is
equivalent to consider all the linearly independent colors composed by the HST
magnitudes and the J-K color to estimate ages, metallicities and reddenings.} 

\begin{eqnarray}
\chi^{2}&=&\sum_{i=1}^{N} \frac{(M_{i}^{mod}+DM+E(B-V)~k_{i}-2.5~\log M -
m_{i}^{obs})^{2}}{\sigma(m_{i})^{2}+\epsilon^{2}}+\nonumber\\
& &\frac{((J-K)^{mod}-(J-K)^{obs})^{2}}{\sigma(J-K)^{2}+2~\epsilon^{2}}
\end{eqnarray}

where $m_{i}^{obs}$ and $M_{i}^{mod}$ are the i-th magnitude and the 
prediction of the stellar population model for 1 $M_{\odot}$ at a given
metallicity and age, $\sigma(m_{i})$ is the error on the magnitude $m_{i}^{obs}$,
$DM=27.41$ is the distance modulus of NGC 4214 (Dalcanton et al. 2009), $k_{i}=A_{\lambda}/E(B-V)$
is the reddening coefficient in the i-th band (from Girardi
et al. 2008, Marigo et al. 2008), $\log M$ is the logarithm of the cluster
mass,
$N$ is the number of adopted magnitudes
and $\epsilon$ is a softening parameter (set to 0.05 mag in all filters; see
Whitmore et al. 2010) which accounts for the models errors.
Note that the logarithm of the cluster mass is included in eq. 1 as a constant term
to be subtracted to each magnitude. So, while ages and metallicities are determined
only by colors, masses are determined from the scaling
factor providing the bestfit magnitudes.
As the apertures adopted for some clusters in the J and K images differ from
those defined by U07b (see Sect. \ref{obs_sec}), we have not used them to constrain 
cluster masses.
Errors have been calculated estimating the parameter interval where the
probability associated to the $\chi^{2}$ include 68\% of the probability
integrated over the entire parameter range\footnote{Note that the errors do
not depend on the absolute value of the $\chi^{2}$ but on its variation across
the parameter space. For this reason a few clusters with relatively large values of
$\chi^{2}/n$ (possibly due to model inadequaces; see Sect. \ref{lmc_sec}) in Table 2 can have small errors 
in the derived parameters.}
In our procedure we adopted only the F170W, F336W and F555W magnitudes derived
by U07b, excluding the F702W and F814W ones. In fact, these magnitudes have only a poor discriminating power and 
for many clusters observations significantly deviate from the model
predictions by several times the nominal uncertainties. This is shown in
Fig. \ref{colex} where the $J-K~vs.~F555W-F702W$ and the $J-K~vs.~F555W-F814W$
color-color diagrams are shown. An age degeneracy is apparent in both planes
during the Red Supergiants phase (at $0.5<J-K<1$) which is noticeable as a closed loop in these
diagrams. Furthermore, the observed F555W-F814W colors stray from the
model prediction by several times their formal error, indicating a possible
inadequacy of models and/or photometric calibration.
It is worth noting that the apparent discrepancy between models and
observations reported in the previous section could in principle affect age 
determinations. In this regard, when the clusters' colors are driven by the
contribution of Main Sequence stars (at ages
$\log{t/yr}<6.8$ and $\log{t/yr}>8$; see Fig. \ref{mult}b) clusters follow a
monotonic ultraviolet-infrared 
color-color relation which is well reproduced by models. On the other hand,
in the age interval 
7$<\log{t/yr}<$7.5 J-K colors are dominated by Red Supergiants and are expected 
to exhibit a significant infrared excess which 
can be easily identified by the $\chi^{2}$ minimization technique described 
above even if the value of the J-K color is not exactly predicted by models.
The inadequacy of the models could instead affect age determinations for
clusters in the
age interval immediately before (6.8$<\log{t/yr}<$7) and after (7.5$<\log{t/yr}<$8) 
the maximum red shift, when the contribution of Main Sequence and Red
Supergiants balance between them.
The ages of these clusters depend on the relative contribution of Main Sequence
and Red Supergiants which is not correctly reproduced by theoretical models. 
As a consequence, for these clusters an intermediate age could be erroneously
assigned. These age ranges are however relatively short and only few clusters
are expected to be affected by this uncertainty.  

In principle, all parameters can be allowed to vary across the parameter space.
However, to limit the degeneracy of our solutions we considered two cases:
{\it i)} constrain 
cluster-to-cluster metallicitiy variations
to lie within a range of $\Delta [M/H]=0.5~dex$ around
a mean value, and {\it ii)} keep metallicity fixed to the spectroscopic measure
by Kobulnicky \& Skillman (1996; [M/H]=-0.5). This is an appropriate choice
as it is not plausible that a larger chemical enrichment could occur in the
recent star formation history of this galaxy (see also W11).
Reddening can instead vary across the field of view by several magnitudes being
very patchy (U07a). Leaving reddening as a free parameter has
however the drawback of favoring the well known age-reddening degeneracy, i.e.
old clusters with intrinsically red colors can be also identified as young
clusters with a high degree of reddening. In fact, following this approach two
clusters have two solutions with similar likelihood (clusters I-Es and IVs) 
appearing
very young and obscured ($\log{t/yr}<7$; $E(B-V)>0.4$) when reddening is left as
a free parameter, while they result significantly older ($\log{t/yr}>8$) when adopting a small 
extinction value.
So, we also performed our analysis
constraining the reddening to vary from cluster-to-cluster within a range of 
$\Delta E(B-V)=0.1$ around a mean value. The mean values of metallicity and
reddening have been chosen as those that minimize the sum of the $\chi^{2}$ of
the 10 clusters.
In Table 2 the results of our analysis are listed for all the above described
assumptions on metallicity and reddening. We note that all the approaches lead to very similar 
estimated ages. The differences between the solutions obtained with the two
assumptions on metallicity are consistent within $\Delta\log{t/yr}<0.3$. When the reddening is constrained to vary by only a small amount
the two clusters with multiple solutions (I-Es and IVs) appears both relatively
old ($\log{t/yr}>8$). The lack of a
significantly intense $H\alpha$ emission in the direction of these clusters
(MacKenty et al. 2000) favors this last hypothesis.

In Fig.s \ref{confrt} and \ref{confrm} the ages and masses of the 10 YMCs derived in this paper
are compared with those estimated by U07a. For a few clusters (IEs, IIIs and 
IVs in U07a; IEs and IVs in our analysis conducted leaving reddening as a free 
parameter) two independent solutions with young and old ages are provided. 
In these cases the comparison has been made between the pair of young and old
solutions obtained by both works. It is immediately evident that  
while the ages we derived for clusters belonging to the star forming 
complex I agree within the errors with those estimated by U07a, 
our ages are on average larger for all clusters belonging to the 
star forming complex II. Some differences between the analysis
performed here and that by U07a could play a role: the adopted stellar models, 
the different resolution of the two datasets, etc. However,
the main reason for the different age estimates stands in the different 
choice of the filters for the $\chi^{2}$ minimization algorithm. 
In particular, in our
work the J-K color accounts for a significant fraction of the entire
likelihood.
Instead, in U07a only few clusters (none of them belonging to complex II) 
have available infrared magnitudes. Moreover, the large errors associated to 
the 2MASS magnitudes
give negligible weights to infrared colors and the age determination is
driven by the ultraviolet-optical properties of the clusters. 
So, the large J-K excess visible in Fig. \ref{colcol} is interpreted in our
analysis as the effect of a significant contribution of Red Supergiants
which are effective after $\sim6~10^{6}~yr$ at $[M/H]\sim0$.

In Fig. \ref{confrebv} the reddening estimates are compared. Of course, a
meaningful comparison is possible only for the estimates made without any
prior. Also in this case, the reddening estimated for clusters belonging to 
complex I agree within the errors with those provided by U07a, while for 
clusters in complex II our estimates are significantly smaller. This is also a
consequence of the age-reddening degeneracy: to reproduce optical colors (which
come from the same analysis of U07a) the effect on colors produced by the 
older ages we derived for these clusters is compensated by a smaller reddening.

Masses appears instead only slightly larger than those listed by U07a. This is an
implicit consequence of the above discussed age difference: since the 
mass-to-light ratios of old clusters are larger than those of young ones,
the same luminosity is interpreted as a larger mass in the former case. 
Reddening differences play also a role: in clusters where significantly 
smaller reddenings have been estimated (see above), the effect on magnitudes 
is compensated by a slightly smaller mass.

\section{The ultraviolet-infrared plane as a tool to identify 
multiple-populations in YMCs?}

\begin{figure*}
 \includegraphics[width=16cm]{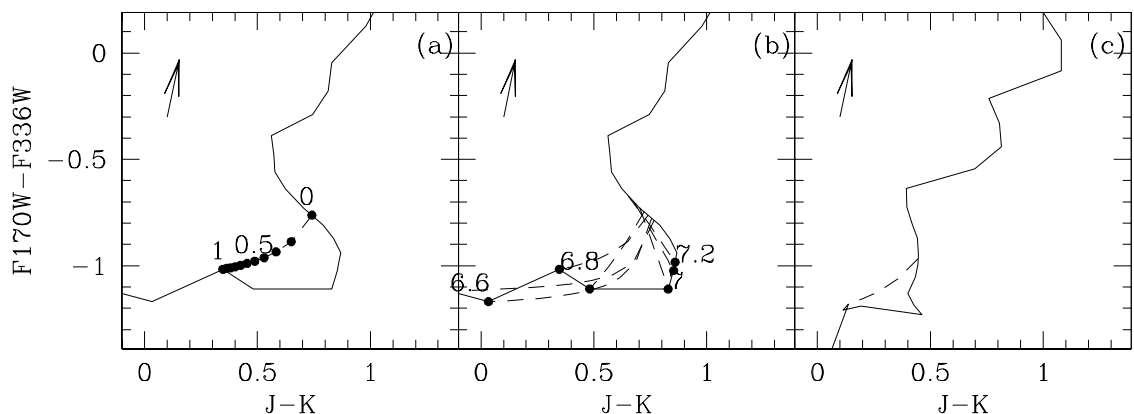}
\caption{Locus of a cluster formed by first/second generations with different second 
generation mass fractions (dashed lines). In panel (a) the ages
$\log{t/yr}=$7.6 and 6.8 for the two generations and a metallicity of 
$[M/H]=0$ have been adopted. In
panel (b) an age differences of $\Delta t=30~Myr$ between the two generations 
and different ages of the second generation have been adopted. 
Panel (c) is the same of panel (a) but for a metallicity
$[M/H]=-1.0$. Evolutionary tracks of single stellar population clusters 
(solid lines) and the reddening vector corresponding to E(B-V)=0.1 are shown 
in all panels.}
\label{mult}
\end{figure*}

The peculiar shape of the ultraviolet-infrared color-color diagram shown in Fig.
\ref{colcol} can
in principle represent an useful tool to select good candidates to host
multiple populations. Indeed, the combination of two stellar populations
with similar metallicities and different ages would produce a cluster with a
color
$$A-B=(A_{f}-B_{f})-2.5~log\left(\frac{1-\eta+\eta~10^{0.4(A_{f}-A_{s})}}
{1-\eta+\eta~10^{0.4(B_{f}-B_{s})}}\right)$$
where A-B is an arbitrary color, $\eta$ is the relative mass fraction of the
second generation and the subscripts refer to
first ($f$) and second ($s$) generations.
In Fig. \ref{mult}a the location in the $(F170W-F336W)~vs.~(J-K)$ color-color
diagram of a cluster formed by two generations of stars with ages
$\log{t/yr}$=7.6 and 6.8 ($\Delta t\sim30~Myr$), [M/H]=0 and different population ratios $\eta$
is compared with the track of single population clusters with the same
metallicity. It is evident that multiple populations clusters with
$0.1<\eta<0.5$ occupy a region forbidden to single population clusters. The identification
of some cluster in this region could therefore indicate the presence of
multiple populations. 
In panel b the same diagram is shown considering different ages for the
second generation (adopting always an age difference between the two
populations of $\Delta t\sim30~Myr$; the minimum evolutionary timescale for a 
massive AGB star; D'Antona \& Ventura 2008). It is evident that clusters cross the 
forbidden region only if the second generation is younger than 
$\log{t/yr}<7$. 
Finally, in panel c the same diagram is shown assuming a
lower metallicity $[M/H]=-1.0$. In this case, the extent of the red 
excursion due to the Red Supergiants is significantly reduced (because of
ombined effect of the higher temperature of the Hayashi track and of the
shorter time spent in the Red Supergiants phase) and the forbidden 
region almost disappears. A very similar conclusion has been reached by Peacock
et al. (2013) who however consider optical colors instead of infrared ones. As
shown in Sect. \ref{uvir_sec}, infrared colors are more suitable for
this purpose, maximizing the extent of the turnover thus allowing a much clear
identification of multiple populations clusters.

Unfortunately, many sources of scatter in this diagram complicate the task.
First, as discussed above, the position of the multiple population region depends on
metallicity. The above method works only when relatively high ($[M/H]>-0.5$) 
and homogeneous metallicities are involved. This is however the
situation for most galaxies, where the most recent episodes of star
formation are characterized by high metal content (Terlevich \& Forbes 2002).
Second, most YMCs are often embedded in dust-rich clouds
with strong absorption up to several visual magnitudes (like in the Antennae
galaxies and M83; Mengel et al. 2005; Kim et al. 2012). 
This has only a minor effect on infrared colors but produces a significant
shift in ultraviolet ones. Hence, the location of multiple population clusters
can be reached also by young but highly reddened single population clusters.
The effects of the above sources of uncertainties can be partly reduced by adopting
several combinations of ultraviolet colors which can break the
age-metallicity-reddening degeneracy.
Third, uncertainties in the calibration and the aperture selection can
introduce a spread of several tenths of magnitudes (in particular in small
clusters).
Fourth, in young clusters integrated fluxes are dominated by red/blue 
Supergiants which are in fast evolutionary stages. Therefore, only few stars
determine the color of a stellar populations and small number fluctuations
can introduce significant scatter in the above diagram (see the discussion in
Gazak et al. 2012). This last effect can
be however negligible if massive ($M>10^{6}M_{\odot}$) clusters are
considered. 
Fifth, as shown in Sect. \ref{lmc_sec}, models often provide a poor
representation of the observational data. This can be due for instance to
the uncertainties in color-temperature transformation (see Fig.
\ref{ngc2100}) or to the uncertainties in the core/envelope overshooting
efficiency which determines the fraction of blue-loop/Red Supergiants stars
(McQuinn et al. 2011). In this regard, the agreement between the 
distribution of clusters in the $(U-V)~vs.~(J-K)$ diagram for both NGC 4214
and the LMC (Fig. \ref{lmc}) seems to suggest the possibility of using {\it empirical}
evolutionary sequences to define the locus of single stellar population
clusters in these diagrams.

\section{Discussion and summary}

In this work we used the infrared JK integrated magnitudes obtained from a dedicated NICS@TNG 
survey of 10 YMCs in the nearby starburst galaxy NGC 4214 to derive their ages, 
metallicities, reddening and masses. 
By combining infrared magnitudes with ultraviolet ones already available from 
high-resolution HST observations we note that these clusters are located in
the combined ultraviolet-infrared color-color planes on well defined sequences, whose
shapes allow a precise determination of their age, with a resolution of $\sigma
(\log{t/yr})\sim0.1$. 
From our analysis all clusters are relatively young ($6.7<\log{t/yr}<8.3$, with
a median value of $\log{t/yr}=7.15$) and 
metal-rich ($-0.5<[M/H]<0$) with masses in the range $4.1<\log{M/M_{\odot}}<5.7$. 
For comparison, the young ($<1~Gyr$) population
of clusters of the LMC has an age distribution peaked at a significantly larger
age ($\log{t/yr}\sim7.9$; Popescu, Hanson \& Elmegreen 2012), as expected given the higher
star-formation rate of NGC 4214 with respect to the LMC. 
While the derived ages of clusters in the 4214-I complex are in good agreement with those 
estimated by U07a, the clusters in the 4214-II complex appear to be significantly older 
than what previously estimated. The difference stems from the extremely red J-K 
color of many clusters, which is interpreted as due to a significant contribution of Red
Supergiants (which are effective after $\sim10^{7}~yr$ at $[M/H]\sim0$).
The detection of a significant population of Red Supergiants in these 
objects through future resolved HST photometry could indicate the 
veridicity of this last interpretation.

The estimated metallicities are in agreement with that provided by
W11 for the young field population of NGC 4214
($-0.6<[M/H]<0$) and are higher than the mean metallicity of the old galactic
population ($-1.6<[M/H]<-0.6$).

It can be noted that even leaving reddening as a free
parameter the derived reddening does not exceed $E(B-V)=0.55$, with a median
value of $E(B-V)=0.065$. On the other hand, if we constrain reddening to lie 
within a 0.1 range, the bestfit range turns out to be $E(B-V)=0.08\pm0.05$. The 
above values are compatible with the estimated foreground Galactic reddening by 
Schlegel, Finkbeiner \& Davis (1998). This indicates that, on average, the YMCs of NGC 4214 
lie in a relatively optically thin region.
Similar conclusions have been reached by Lee et al. (2009) and W11. 
This also explain the well defined turnover in the ultraviolet-infrared 
colors.

The masses of these clusters are only slightly smaller than those of present-day 
Galactic globular clusters. The most
massive and compact among these objects will evolve toward a globular
cluster-like structure. It is therefore possible to
consider these objects as the young counterparts of globulars. 
Note however that, according to the scenario proposed by D'Ercole et al.
(2008), Galactic globular clusters were $\sim$10 times more massive than the 
YMCs of NGC 4214. It is therefore unprobable that these clusters will undergo the
same chemical evolution of Galactic globular clusters.

We discussed the possibility of using ultraviolet-infrared color-color
diagrams to test the presence of multiple populations in YMCs. While we find
no evidence of multiple populations in the YMCs of NGC 4214, this tool can
be in principle effective in selecting good candidates for clusters
characterized by high-metallicity ($[M/H]>-1$), low reddening ($E(B-V)<0.1$) and 
a young ($\log{t/yr}<7$) second generation 
constituting a cluster mass fraction in the range $0.1<\eta<0.5$. 
Although the above constraints reduce the detection efficiency and the
inadequacy of stellar evolution models in infrared bands prevents from any
conclusive detection, the 
ultraviolet-infrared color-color diagrams can be used in statistical sense if
many YMCs are observed. Galaxies
characterized by a high star formation rate (e.g. starburst galaxies)
represent a good benchmark to test this tool providing a large number of
YMCs within a relatively chemically homogeneous environment.

\section*{Acknowledgments}

We thank the anonymous referee for his/her useful comments and suggestions.
A.S. acknowledge the support of INAF through the 2010
postdoctoral fellowship grant. AB, RG, EC, and SL acknowledge the PRIN INAF 2009 "Formation and Early
Evolution of Massive star Clusters" (PI R. Gratton); AS, EC, SL
acknowledge the PRIN INAF 2011 "Multiple populations in globular
clusters: their role in the Galaxy assembly" (PI E. Carretta); and AB,
EC, and SL aknowledge the PRIN MIUR 2010-2011 ``The Chemical
and Dynamical Evolution of the Milky Way and Local Group Galaxies'' (PI
F. Matteucci), prot. 2010LY5N2T


\label{lastpage}

\end{document}